\documentclass[twocolumn]{aastex6}
\usepackage{graphicx}
\usepackage{textcomp}

\newcommand{\sunrise}{\textsc{Sunrise}}

\begin{document}
%opening
\title{Estimation of the magnetic flux emergence rate in the quiet Sun from \sunrise{} 
data}
\author{\textsc{H.~N.~Smitha,$^{1}$ 
L.~S.~Anusha,$^{1}$ 
S.~K.~Solanki,$^{1,2}$ and 
T.~L.~Riethm\"{u}ller$^{1}$
}}
\affil{
$^{1}$Max-Planck-Institut f\"ur Sonnensystemforschung, Justus-von-Liebig-Weg 3, 37077 
G\"ottingen, Germany\\ smitha@mps.mpg.de\\
$^{2}$School of Space Research, Kyung Hee University, Yongin, Gyeonggi, 446-701, Republic 
of Korea\\
}

\begin{abstract}
Small-scale internetwork (IN) features are thought to be the major source of fresh 
magnetic flux in the quiet Sun. During its first science flight in 2009, the balloon-borne 
observatory  \sunrise{} captured images of the magnetic fields in the quiet Sun at a high 
spatial resolution. Using these data we measure the rate at which the IN features bring 
magnetic flux to the solar surface.  {In a previous paper it was found that the 
lowest magnetic flux in small-scale features detected using the} \sunrise{} 
 {observations is $9\times10^{14}$ Mx.} This is nearly an order of magnitude smaller 
than the smallest fluxes of features detected in observations from \textit{Hinode} 
satellite.  {In this paper, we compute the flux emergence rate (FER) by accounting 
for such small fluxes, which was not possible before} \sunrise. By tracking the features 
with fluxes in the range $10^{15}-10^{18}$ Mx, we measure an FER of 
$1100\rm{\,\,Mx\,cm^{-2}\,day^{-1}}$. The smaller features with fluxes $\le 10^{16}$\,Mx 
are found to be the dominant contributors to the solar magnetic flux. The FER found here 
is an order of magnitude higher than the rate from the \textit{Hinode}, obtained with a 
similar feature tracking technique. A wider comparison with the literature shows, 
however, that the exact technique of determining the rate of the appearance of new flux 
can lead to results that differ by up to two orders of magnitude, even when applied to 
similar data. {The causes of this discrepancy are discussed and first qualitative 
explanations proposed.}
\end{abstract}

\keywords{Sun: atmosphere, Sun: magnetic fields, Sun: photosphere}

\maketitle

\section{Introduction}
\label{s1}
The quiet Sun covers most of the solar surface, in particular at activity minimum, but 
also plays an important role even during the active phase of the solar cycle. The magnetic 
field in the quiet Sun is composed of the network \citep{1967SoPh....1..171S}, 
internetwork \citep[IN,][]{1971IAUS...43...51L, 1975BAAS....7..346L}, and the ephemeral 
regions \citep{1973SoPh...32..389H}. For an overview of the small-scale magnetic 
features, see \citet{1993SSRv...63....1S, 2009SSRv..144..275D, 2014masu.book.....P, 
2014A&ARv..22...78W}.

The IN features are observed within the supergranular cells and carry hecto-Gauss fields  
\citep[][]{1996A&A...310L..33S, 2003A&A...408.1115K, 2008A&A...477..953M} although 
kilo-Gauss fields have also been observed in the IN  \citep{2010ApJ...723L.164L, 
Lagg16}. They evolve as unipolar and bipolar features with typical lifetimes of less than 
10 minutes \citep{2010SoPh..267...63Z, 2013apj...774..127l,lsa}, i.e., they continuously 
bring new flux to the solar surface, either flux that has been either freshly generated, 
or recycled. They carry fluxes $\le 10^{18}$ Mx, with the lower limit on the smallest 
flux decreasing with the increasing spatial resolution and polarimetric sensitivity of 
 the observing instruments, although the identification technique also plays an important 
role. 

Ephemeral regions are bipolar magnetic features appearing within the supergranular cells 
carrying fluxes $\approx 10^{19}$ Mx \citep{2001ApJ...548..497C, 2003ApJ...584.1107H} and 
are much longer-lived compared to the IN features, with lifetimes of 3 -- 4.4 hours 
\citep{2000RSPTA.358..657T, 2001ApJ...555..448H}. The ephemeral regions also bring new 
magnetic flux to the solar surface. 

The network is more stable, with typical lifetimes of its structure of a few 
hours 
to a day, although the individual kG magnetic elements within the network live for a much 
shorter time, as the entire flux within the network is exchanged within a period of 8--24 
hr \citep{2003ApJ...584.1107H, 2014apj...797...49g}. The flux in the network is fed by 
ephemeral regions \citep{0004-637x-487-1-424, 2001ApJ...555..448H} and IN features 
\citep[][]{2014apj...797...49g}. The network features are found along the supergranular 
boundaries and carry fields of kG strength with a typical flux of $10^{18}$ Mx 
\citep{1995soph..160..277w}. 

\begin{figure*}
\centering
\includegraphics[width=0.7\textwidth]{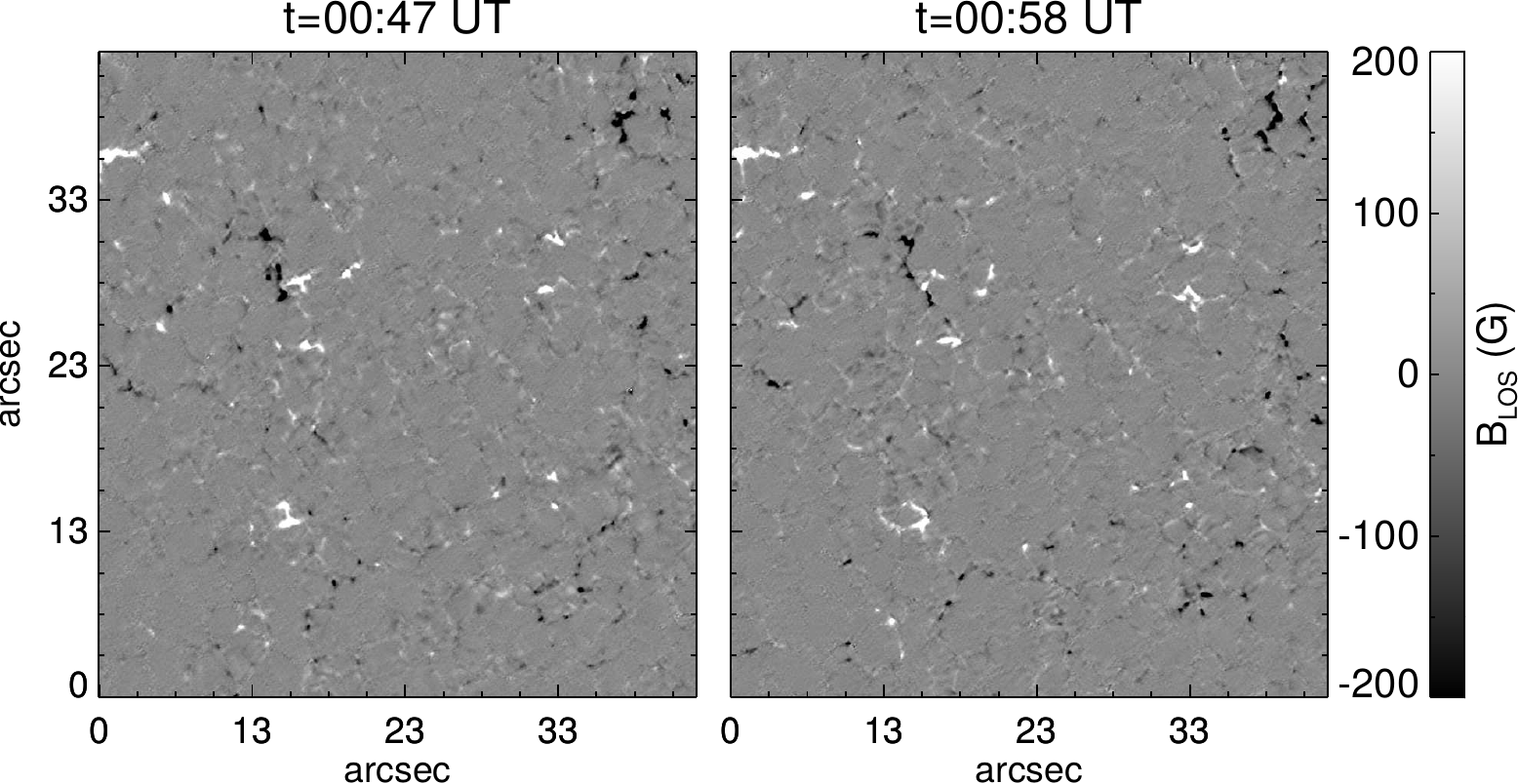}
\caption{Two sample magnetograms recorded at $t$ = 00:47 UT and 00:58 UT on 2009 June 9 
with the \sunrise/IMaX instrument during its first science flight.}
\label{magnetogram}
\end{figure*}

The magnetic flux is produced by a dynamo, the location of which is currently the subject 
of debate, as is whether there is only a single dynamo acting in the Sun 
\citep[e.g.,][]{2012A&A...547A..93S} or whether there is a small-scale dynamo acting in 
addition to a global dynamo \citep{1993A&A...274..543P, 1999ApJ...515L..39C, 
2001A&G....42c..18C, 2007A&A...465L..43V, 2008A&A...481L...5S, 2010A&A...513A...1D, 
2013A&A...555A..33B, 2015ApJ...803...42H, 2016ApJ...816...28K}. In addition, it is unclear 
if all the magnetic flux appearing on the Sun is actually new flux produced by a dynamo, 
or possibly recycled flux transported under the surface to a new location, where it 
appears again \citep[e.g.,][]{2001ASPC..236..363P}. This may be particularly important at 
the smallest scales. 

An important parameter constraining the production of magnetic flux is the amount of 
magnetic flux appearing at the solar surface. In particular, the emergence of magnetic 
flux at very small scales in the quiet Sun provides a probe for a possible small-scale 
dynamo acting at or not very far below the solar surface. The deep minimum between solar 
cycles 23 and 24 offered a particularly good chance to study such flux emergence, as the 
long absence of almost any activity would suggest that most of the emerging flux is newly 
produced one and is not flux transported from decaying active regions to the quiet Sun 
(although the recycling of some flux from ephemeral regions cannot be ruled out).

The IN quiet Sun displays by far the largest magnetic flux emergence rate (FER). Already, 
\citet{1987SoPh..110..101Z} pointed out that two orders of magnitude more flux appears in 
ephemeral regions than in active regions, while the FER in the IN is another two orders 
of 
magnitude larger. This result is supported by more recent studies 
\citep[e.g.,][]{2002ApJ...565.1323S, 2009SSRv..144..275D, 2009ApJ...698...75P, 
2011SoPh..269...13T}.  Given the huge emergence rate of the magnetic flux in the IN, it 
is of prime importance to measure the amount of flux that is brought to the surface 
by these features.
 
The current estimates of the FER in the IN vary over a wide range, which include: 
$10^{24}\rm{\,Mx\, day^{-1}}$ \citep[][]{1987SoPh..110..101Z}, $3.7\times 
10^{24}\rm{\,Mx\,day^{-1}}$ 
\citep[$120\rm{\,Mx\,cm^{-2}\,day^{-1}}$,][]{2016ApJ...820...35G} and 
$3.8\times10^{26}\rm{\,Mx\,day^{-1}}$ \citep[][]{2013SoPh..283..273Z}.
By considering all the magnetic features (small-scale features and active regions), 
\citet{2011SoPh..269...13T} measure a global FER of $3\times 10^{25}\rm{\,Mx\,day^{-1}}$ 
($450 \rm{\,Mx \,cm^{-2} \,day^{-1}}$), while \citet{th-thesis} measures $3.9 \times 
10^{24} \rm{\,Mx \,day^{-1}}$ ($64\rm{\,\,Mx\,cm^{-2}\,day^{-1}}$), whereby almost all of 
this flux emerged in the form of small IN magnetic features. The FER depends on the 
observations and the method used to measure it. A detailed comparison of the FERs from 
different works is presented Section~\ref{s6}.

To estimate the FER, \citet{1987SoPh..110..101Z} and \citet{2011SoPh..269...13T}  
considered features with fluxes $\ge 10^{16}$ Mx, while \citet{2013SoPh..283..273Z} and 
\citet{2016ApJ...820...35G} included features with fluxes as low as $6\times 10^{15}$ 
Mx and $6.5\times10^{15}$ Mx (M. Go\v{s}i\'{c}, priv. comm.), respectively. However, with 
the launch of the balloon-borne \sunrise{} observatory in 2009 \citep{2010ApJ...723L.127S, 
2011SoPh..268....1B, 2011SoPh..268..103B, 2011SoPh..268...35G} carrying the Imaging 
Magnetograph eXperiment \citep[IMaX,][]{2011SoPh..268...57M}, it has now become possible 
to estimate the FER including the contribution of IN features with fluxes as low as 
$9\times10^{14}$ Mx \citep[][hereafter referred to as LSA17]{lsa}. The IMaX instrument 
has 
provided unprecedented high-resolution magnetograms of the quiet Sun observed at 5250\, 
\AA. The high resolution is the main reason for the lower limiting flux. A detailed 
statistical analysis of the IN features observed in Stokes $V$ recorded by \sunrise/IMaX 
is carried out in LSA17. In the present paper we estimate the FER in the IN region using 
the same data.

In Section~\ref{s2} we briefly describe the employed \sunrise{} data. The IN features 
bringing flux to the solar surface that are considered in the estimation of FER are 
outlined in Section~\ref{s3}. The FER from \sunrise{} are presented, discussed and 
compared with previously obtained results in Section~\ref{s4} while our conclusions 
are presented in Section~\ref{s8}.

\begin{figure*}
\centering
\includegraphics[width=0.9\textwidth]{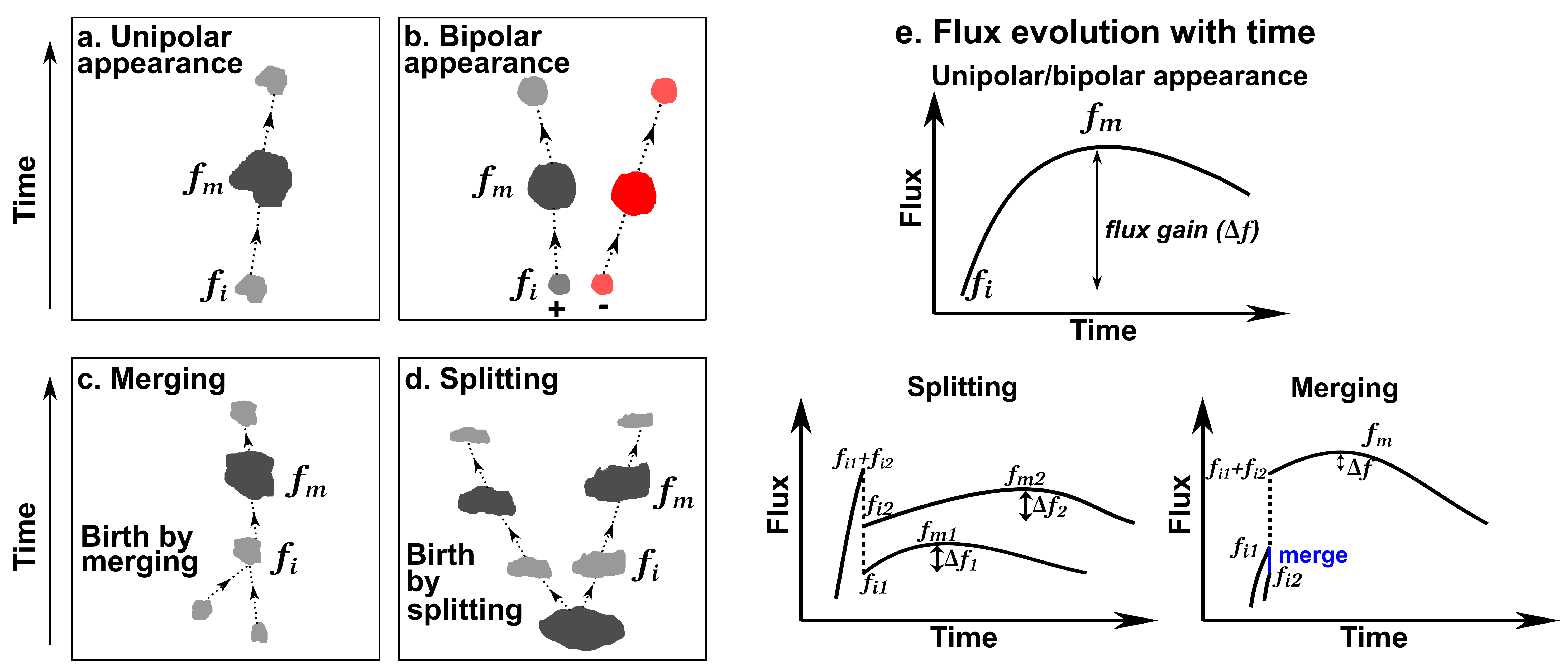}
\caption{Panels a--d: Schematic representation of the different paths by 
which magnetic flux is brought to the solar surface and of its subsequent evolution.  The 
quantities $f_i$ and $f_m$ are the instantaneous and maximum fluxes of a feature, 
respectively, where instantaneous means just after it appears. {Panel e}: Typical 
variation in the flux of a feature, born by  {unipolar and bipolar} appearances 
(top) and born by splitting/merging (bottom), over time. The flux gained by them after 
birth is $\Delta f=f_m - f_i$. The features born by splitting carry fluxes $f_{i1,2}$ at 
birth and reach $f_{m1,2}$ in the course of their lifetime, gaining flux $\Delta f_{1,2}$ 
after birth. $f_{i1}+f_{i2}$ is equal to the flux of the parent feature at the time of its 
splitting. The feature born by merging carries flux equal to the sum of the fluxes of the 
parent features $f_{i1}+f_{i2}$. The blue line indicates the merging of the two features.}
\label{flux}
\end{figure*}

\section{Data}
\label{s2}
The data used here were obtained during the first science flight of \sunrise{} described 
by \citet{2010ApJ...723L.127S}. We consider 42 maps of the line-of-sight (LOS) magnetic 
field, $B_{\rm LOS}$, obtained from sets of images in the four Stokes parameters  
recorded with the IMaX instrument between 00:36 and 00:59 UT on 2009 June 9 at the solar 
disk center, with a cadence of 33\,s,  {a spatial resolution of 
$0.^{\prime\prime}15-0.^{\prime\prime}18$ (plate scale is $0.^{\prime\prime}054$ per 
pixel), and an effective field of view (FOV) of $43^{\prime\prime} \times 
43^{\prime\prime}$ after phase diversity reconstruction}.  {The data were 
reconstructed with a point spread function determined by in-flight phase diversity 
measurements to correct for the low-order aberrations of the telescope \citep[defocus, 
coma, astigmatism, etc., see][]{2011SoPh..268...57M}. The instrumental noise of the 
reconstructed data was $3\times10^{-3}$ in units of continuum intensity. For 
identification of the features, spectral averaging was done which further reduced the 
noise to $\sigma=1.5\times10^{-3}$. All features with signals above a $2\sigma$ 
threshold, which corresponds to 12\,G,  were used \citep[][]{2011SoPh..268...57M}.}

To measure the flux, we use $B_{\rm LOS}$ determined with the center-of-gravity (COG) 
method 
\citep[][LSA17]{1979A&A....74....1R, 2010A&A...518A...2O}.  {The inclination of the 
IN fields has been under debate, with several studies dedicated to measuring their 
angular distribution. By analysing the data from \textit{Hinode}, 
\citet{2007ApJ...670L..61O, 2008ApJ...672.1237L} concluded the IN fields to be 
predominantly horizontal. However, using the same dataset, \citet{2011ApJ...735...74I} 
found some of the IN fields to be vertical. \citet{2014A&A...569A.105J} arrived at similar 
conclusions (vertical inclination) by analysing the magnetic bright points observed from 
the first flight of \sunrise{}. Variations in the inclination of the IN fields with 
heliocentric angle ($\mu$) have been reported by \citet{2012ApJ...751....2O, 
2013A&A...550A..98B, 2013A&A...555A.132S}. Isotropic and quasi-isotropic distribution of 
the IN field inclinations is favoured by \citet{2008A&A...479..229M}, using the Fe\,{\sc 
i} $1.56\,\mu$m infrared lines, and by \citet{2009ApJ...701.1032A,2014A&A...572A..98A} 
using \textit{Hinode} data. More recently, \citet{2016A&A...593A..93D} found the 
distribution of IN field inclination to be quasi-isotropic by applying 2D inversions on 
\textit{Hinode} data and comparing them with 3D magnetohydrodynamic simulations. For a 
detailed review on this, see \citet{2015SSRv..tmp..113B}.}

 {In the determination of the FER, we use $B_{\rm LOS}$ for consistency and for 
easier comparison with earlier studies on the FER. Also, the determination of the exact 
amount of flux carried in horizontal field features is non-trivial and requires estimates 
of the vertical thickness of these features and the variation of their field strength with 
height. In addition, if they are loop-like structures \citep[as is suggested by 
local-dynamo simulations, e.g.,][]{2007A&A...465L..43V}, then there is the danger of 
counting the flux multiple times if one or both of their footpoints happen to be resolved 
by the \sunrise data. We avoid this by considering only the vertical component of the 
magnetic field. It is likely that we miss the flux carried by unresolved magnetic loops by 
concentrating on Stokes $V$, but this problem is suffered by all previous studies of FER 
and should decrease as the spatial resolution of the observations is increased. With the 
\sunrise{} I data analyzed here having the highest resolution, we expect them to see more 
of the flux in the footpoints of the very small-scale loops that appear as horizontal 
fields in \textit{Hinode} and \sunrise{} data \citep{2010A&A...513A...1D}.}

The small-scale features were identified and tracked using the feature tracking code 
developed in LSA17.  {For the sake of completeness we summarize the most 
relevant results from LSA17 as follows. All the features covering at least 5 pixels 
were considered with Stokes $V$ being larger than $2\sigma$ in each pixel. To determine 
the flux per feature,  the $B_{\rm LOS}$ averaged over the feature, denoted as 
\textlangle$B_{\rm LOS}$\textrangle{} was used.} \textlangle$B_{\rm LOS}$\textrangle{} 
had values up to 200\,G, even when the maximum field strength in the core of the feature 
reached kG values.  {A total of 50,255 features of both polarities were identified. 
The sizes of the features varied from 5-1,585 pixels, corresponding to an area range of 
$\approx 8\times 10^{-3}-2.5$\,Mm$^2$. The tracked features had lifetimes ranging from 
0.55 to 13.2 minutes.} The smallest detected flux of a feature was $9\times10^{14}$ Mx 
and 
the largest $2.5\times10^{18}$ Mx.

At the time of the flight of \sunrise{} in 2009 the Sun was extremely quiet, with no signs 
of activity on the solar disk. Two sample magnetograms at  00:47 UT and 00:58 UT are shown 
in Figure~\ref{magnetogram}. Most of the features in these maps are part of the IN and in 
this paper, we determine the rate at which they bring flux to the solar surface. 

\section{Processes increasing magnetic flux at the solar surface}
\label{s3}
The different processes increasing the magnetic flux at the solar surface are 
schematically represented in Figure~\ref{flux}a -- \ref{flux}d. In the figure, $f_i$ 
refers to the flux of the feature at its birth and $f_m$ is the maximum flux that a 
feature attains over its lifetime. A typical evolution of the flux of a feature born by
 {unipolar or bipolar} appearances, and by splitting/merging is shown in 
Figure~\ref{flux}e (top and bottom, respectively). The gain in the flux of a feature 
after its birth is $f_m-f_i$. Magnetic flux at the surface increases through the 
following processes:

    \begin{table*}
          \centering
          \caption{The instantaneous and maximum fluxes of the features, 
  {integrated over all features and time frames}, in different 
 processes measured in LSA17}
            \begin{tabular}{|c|c|c|c|c|}
            \hline
 Process & Instantaneous flux & Maximum flux & Flux gain & Factor of increase\\
 & ($f_i$ in Mx) & ($f_m$ in Mx) & ($\Delta f= f_m-f_i$ in Mx) & ($f_m/f_i$)\\
 \hline
   {Unipolar appearance} & $4.69\times10^{19}$ & $9.69\times10^{19}$ &  
$4.99\times10^{19}$ &2.06\\
  Splitting & $1.76\times10^{20}$& $2.12\times10^{20}$&$3.60\times10^{19}$&1.20\\
  Merging & $1.64\times 10^{20}$& $1.85\times 10^{20}$&$2.20\times10^{19}$&1.31\\
   {Bipolar appearance} & $3.85\times10^{18}$ & $9.53\times10^{18}$& 
$5.67\times10^{18}$&2.47\\
  \hline
            \end{tabular}
            \label{table1}
        \end{table*}

\begin{enumerate}
 \item{ {Unipolar appearance}}: The birth of an isolated feature with no spatial 
overlap with  any of the existing features in the current and/or previous time frame 
(Figure~\ref{flux}a). 
 \item { {Bipolar appearance}}: Birth of bipolar features, with the two polarities 
closely spaced, and either appearing simultaneously or separated by a couple of time 
frames ( {referred to as} time symmetric and asymmetric emergence in LSA17; 
see also Figure~\ref{flux}b). 
\item { {Flux gained by features in the course of their 
lifetime}}: The gain in the 
flux of a feature in the course of its lifetime, i.e. the increase in flux between its 
birth and the time it reaches its maximum flux, before dying in one way or another, either 
by interacting with another feature, or by disappearing.

 This gain can take place in features born in different ways, be it by growth, or 
through the merging or splitting of pre-existing features (Figures~\ref{flux}c, 
\ref{flux}d and \ref{flux}e).
\end{enumerate}

 {Note that the bipolar appearance of magnetic flux is often referred to as 
`emergence' in earlier papers including LSA17. However, the term `emergence' in FER 
describes the appearance of new flux at the solar surface from all the three processes 
described above. To avoid confusion, we refer to the emergence of  bipolar features as 
bipolar appearance.}  {Of all the newly born features over the entire time series, 
19,056 features were unique (for area ratio 10:1, see Section~\ref{s4}). Among them 48\% 
(8728 features) were unipolar and 2\% (365 features) were part of bipolar appearances. 
Features born by splitting constituted 38\% (6718 features), and 12\% (2226 features) 
were born by merging. The remaining 1019 features correspond to those alive in the 
first frame. A comparison of the rates of birth and death of the features by various 
processes for different area ratio criteria is given in Table~2 of LSA17.}

In the FER estimations, the flux brought by the features born by 
 {unipolar and bipolar} appearances is the maximum flux that they attain 
($f_m$) over their lifetime. In the case of features born by splitting or merging, the 
flux gained after birth is taken as the flux brought by them to the surface. This gain is 
the difference between their flux at birth $f_i$ and the maximum flux they attain $f_m$, 
i.e. $f_m - f_i$. 

\begin{figure}[h!]
\centering
 \includegraphics[width=0.3\textwidth]{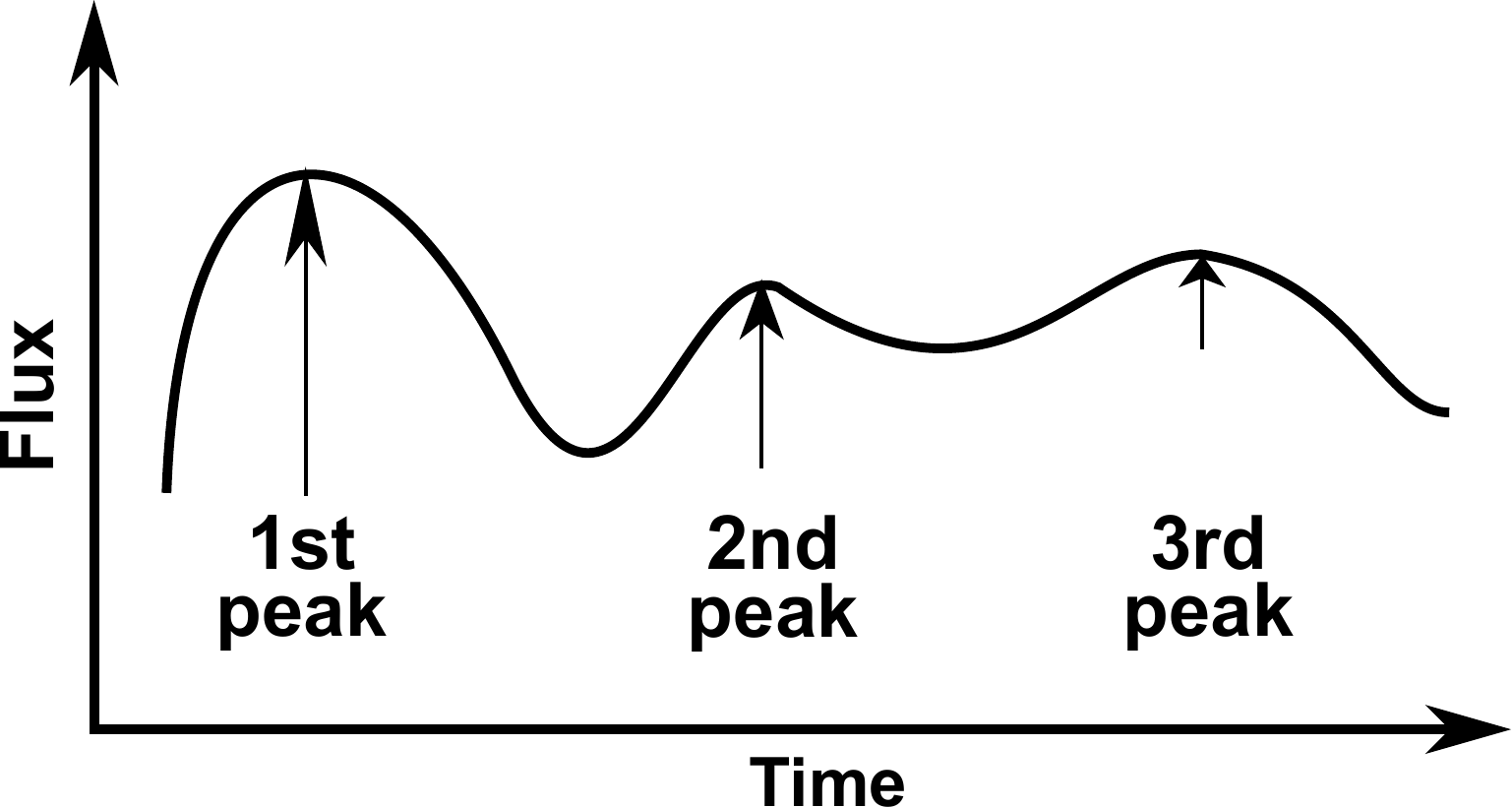}
 \caption{Schematic representation of multiple peaks in the flux of a feature occurring 
in the course of its lifetime. In the FER estimations, we consider only the largest gain, 
i.e., flux increase during first peak in this example.}
 \label{fg}
\end{figure}

Our approach is conservative in the sense that if a feature reaches multiple peaks of flux 
in the course of its lifetime, as in the example shown in Figure~\ref{fg}, then we 
consider only 
the largest one (the flux gained during the first peak in Figure~\ref{fg}), and neglect 
increases in flux contributing to smaller peaks such as the second and third peaks in 
Figure~\ref{fg}.  {Multiple peaks in the flux of a feature (shown in 
Figure~\ref{fg}) are rarely seen, as most features do not live long enough to display 
them (see LSA17). }

 {Changes in the flux of a feature in the course of its lifetime can cause 
it to seemingly appear and disappear with time if its total flux is close to the 
threshold set in the study (given by a signal level twice the noise in at least five
contiguous pixels). If it disappears and reappears again, then it will be counted twice. 
This introduces uncertainties in the measurement of FER. Uncertainties are discussed in 
Section~\ref{s7}.}
% 
% 
% can be 
% observed if it lives long enough. Studies presented in LSA17 indicate that a 
% majority of the features are short lived and especially, the high flux long-lived 
% features are very few in number.  Thus, not accounting for the smaller increases in flux 
% will not significantly affect the FER estimations.}
% 
%  {Also, threshold levels set on the signal level ($2\sigma$) and on area of the 
% features (extending over at least 5 pixels) can cause appearance and disappearance of the 
% magnetic features that are close to the threshold. }

\section{Results and Discussions}
\label{s4}

\subsection{Flux emergence rate (FER)}
\label{s5}
In this work, we consider the results from the area ratio criterion 10:1 of LSA17. In 
that paper, the authors devise area ratio criteria (10:1, 5:1, 3:1 and 2:1) to avoid that 
a feature dies each and every time that a tiny feature breaks off, or merges with it. 
For example in a splitting event, the largest of the features formed by splitting 
must have an area less than $n$ times the area of the second largest, under the $n:1$ area 
ratio criterion.  {We have verified that the choice of the area ratio criterion does 
not drastically alter the estimated FER, with variations being less than 10\% for area 
ratios varying between 10:1 and 2:1.}

The  {instantaneous} and maximum fluxes of the features in different processes are 
given in Tables~$1-5$ of LSA17. A summary is repeated in Table~\ref{table1} for 
convenience {, where fluxes are given for features born by the four processes listed 
in the first column. The instantaneous flux, in the second column, refers to the flux of 
a feature at its birth ($f_i$ in Figure~\ref{flux}). In the third column is the maximum 
flux of a feature during its lifetime ($f_m$ in Figure~\ref{flux}). The flux gain in the 
fourth column is the difference of the second and third columns ($\Delta f=f_m-f_i$). 
In the fifth column is the factor by which the flux increases from its birth to its peak 
($f_m/f_i$). The fluxes given in this table are the sum total over all features in the 
entire time series, for each process.} 

\begin{figure}
\centering
\includegraphics[scale=0.3]{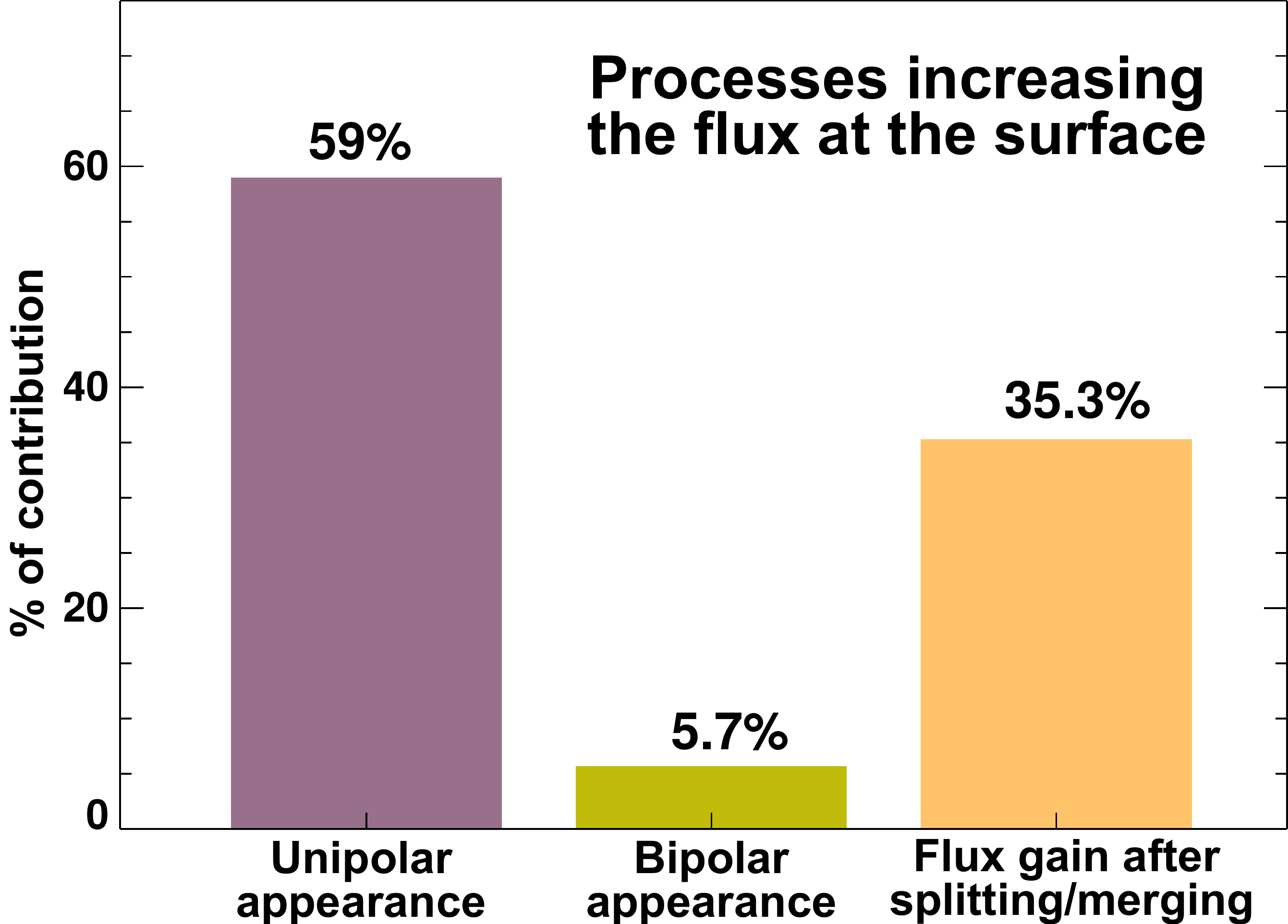}
 %\iplotone{figures/bar}
 \caption{The percentage of contribution to the flux emergence rate (FER) from different 
processes bringing flux to the solar surface. In the case of  {unipolar and bipolar} 
appearances, the maximum flux of the feature is used to determine the FER. For the 
features born by splitting/merging, the flux gained by them after birth is considered. 
This gain is the difference $f_m - f_i$ in Figures~\ref{flux}c, \ref{flux}d and 
\ref{flux}e.}
 \label{flux1}
\end{figure}

To compute the FER, we add the fluxes from the 
various processes described in the previous section. For features born by appearance 
 {(unipolar and bipolar), we take their maximum flux of the features ($f_m$)} to be 
the fresh flux emerging at the surface. For the features formed by merging or splitting 
only the flux increase after the birth  {($\Delta f$)} is considered. From the  
first two processes alone, the total flux brought to the surface is $1.1\times10^{20}$ Mx 
over an FOV of $43^{\prime\prime} \times 43^{\prime\prime}$ in 22.5 minutes. This gives 
an 
FER of $700 \rm{\, \, Mx\, cm^{-2}\,day^{-1}}$. Including the flux gained by split/merged 
features increases the FER to $1100 \rm{\,\,Mx\,cm^{-2}\,day^{-1}}$. Figure~\ref{flux1} 
shows the contribution from each process to the total FER. The isolated features appearing 
on the solar surface contribute the largest, nearly 60\%. Given that the emerging bipoles 
contain only 2\% of the total observed flux (Table~5 in LSA17), they contribute only 
about 
5.7\% to the FER. 

However, the flux brought to the solar surface by features born by splitting or merging, 
after their birth is $5\times10^{19}$ Mx, which is quite significant and contributes 
$\approx35\%$ to the FER. The contribution to solar surface flux by this process is 
comparable to the flux brought to the surface by features born by  {unipolar} 
appearance ($9.7\times10^{19}$ Mx) and nearly an order of magnitude higher than that flux 
from features born by  {bipolar appearance} ($9.5\times10^{18}$ Mx). 

Over their lifetimes, the features born by splitting and by merging gain 1.2 times their 
initial flux (i.e., $f_m = 1.2 \times f_i$). The fluxes gained by features born by 
appearance relative to their flux at birth is slightly higher ($\approx2$ times, i.e., 
$f_m = 2\times f_i$). This is because the initial magnetic flux of the features born by 
appearance is quite low.  The flux at birth of split or merged features is already quite 
high because the parent features which undergo splitting or merging are at later stages in 
their lives (see Figure~\ref{flux}e). This is also evident from the fact that the average 
initial flux per feature of the features born by splitting or merging ($2.9\times10^{16}$ 
Mx and $7.4\times10^{16}$ Mx, respectively) is an order of magnitude higher than the 
average initial flux per feature of the appeared  {unipolar or bipolar} features 
($5.4\times10^{15}$ Mx, see Table~2 of LSA17).

\begin{figure*}
\centering
\includegraphics[width=\textwidth]{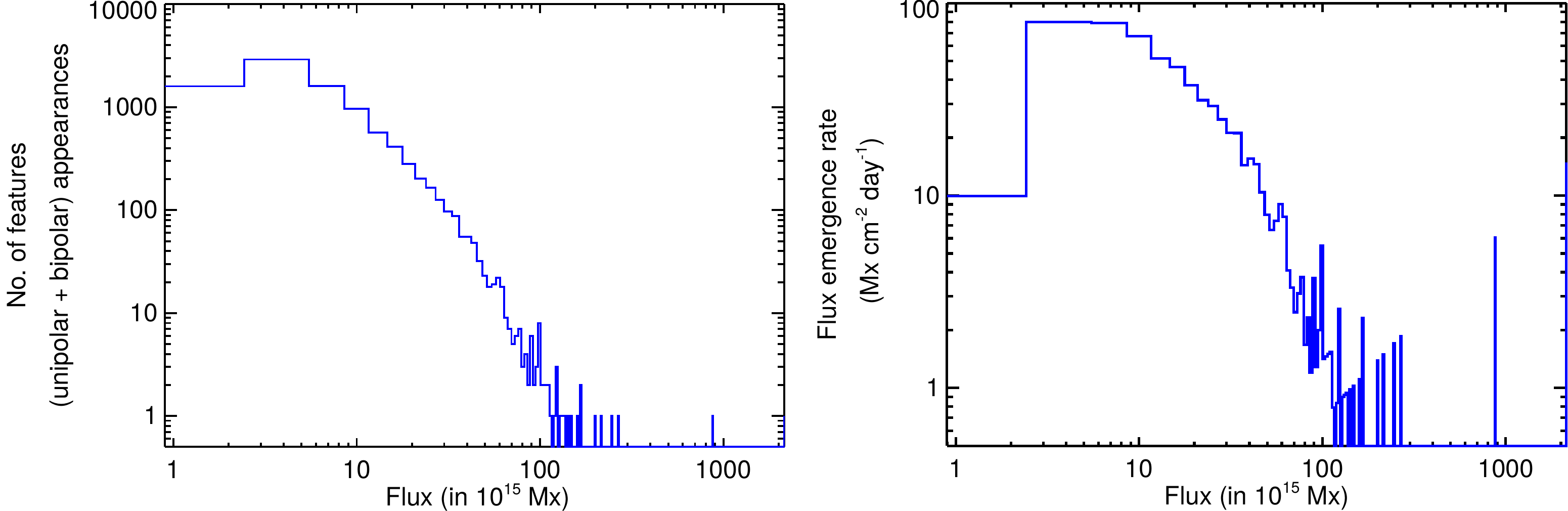}
 \caption{{Left panel:} histogram of the number of features born by 
 {unipolar and bipolar} appearances, carrying fluxes in the range $10^{15}-10^{18}$ 
Mx. {Right panel:} the flux emergence rate from the features born by 
 {unipolar and bipolar} appearances as a function of their flux.}
 \label{hist}
\end{figure*}

The fact that the small-scale magnetic features are the dominant source of fresh flux in 
the quiet photosphere is discussed in several publications \citep{2002ApJ...565.1323S, 
2009SSRv..144..275D, 2009ApJ...698...75P, 2011SoPh..269...13T}. Our results extend these 
earlier findings to lower flux per feature values. As shown in Figure~\ref{hist}, over the 
range $10^{15}-10^{18}$ Mx, nearly $65\%$ of the detected features carry a flux $ \le 
10^{16}$ Mx (left panel). They are also the dominant contributors to the FER (right 
panel). In this figure, only the features that are born by  {unipolar and bipolar} 
appearances are considered. Below $2\times10^{15}$ Mx, we see a drop as we approach the 
sensitivity limit of the instrument. 

\subsubsection{Flux loss rate}
Flux is lost from the solar surface by disappearance, cancellation of opposite 
polarity features, and decrease in the flux of the features in the course of their 
evolution (i.e. the opposite process to the ``flux gain'' described earlier in 
Section~\ref{s3}). As seen from Tables~3 and 4 of LSA17, the increase in flux at the 
solar surface balances the loss of flux, as it obviously must if the total amount of flux 
is to remain unchanged. To compute the flux loss rate, we take the maximum flux of the 
features that die by cancellation and by disappearance to be the flux lost by them in the 
course of their lifetime and during disappearance or cancellation. For the 
features that die by splitting/merging we take the difference between the maximum flux of 
the features and the flux at their death as a measure of the flux lost during their 
lifetimes. By repeating the analyses for the 10:1 area ratio criterion, we find that the 
flux is lost from the solar surface at a rate of 1150$\rm{\, \, Mx\, cm^{-2}\,day^{-1}}$ 
which corresponds within 4.5\% to the obtained FER. This agreement serves as a 
consistency check of the FER value that we find.

\subsection{Uncertainties}
\label{s7}
 {Although most of the uncertainties and ambiguities that arise during feature 
tracking have been carefully taken care of, as discussed in LSA17, some additional 
ones which can affect the estimated FER are addressed below.}

 {In our computation of the FER, the features born before the time series began and 
the features still alive at the end are not considered. According to LSA17, the first and 
the last frames of the time series had 1019 and 1277 features, respectively.  To estimate 
their contribution, we assume that the features still living at the end have a similar 
lifetime, size, flux distribution and formation mechanism as the total number of features 
studied. We attribute the appropriate  average flux at birth and the average flux gain 
for 
features born by splitting, merging, unipolar and bipolar appearance. After including 
these additional fluxes, we get an FER of $\approx 1150 \rm{\, \, 
Mx\,cm^{-2}\,day^{-1}}$, 
corresponding to a $4-5\%$ increase. With this method, we are associating the features 
with flux gain than they might actually contribute (as many of them are likely to reach 
their maximum flux only after the end of the time series). This will be balanced out by 
not considering the features that are already alive at the beginning (also, it is 
impossible to determine the birth mechanism of these features).}

 {Furthermore, in the analysis of LSA17, the features touching the spatial 
boundaries were not counted. An estimate of their contribution, in ways similar to the 
above, leads to a further increase of the FER by $5-6\%$. Thus combining the features in 
the first and last frames and the features touching spatial boundaries together increase 
the FER by $\approx 10\%$. } 

Meanwhile, as discussed in Section~\ref{s3}, in the case of flux gained after birth by 
features born from splitting or merging, we consider only the gain to reach the maximum 
flux in the feature and not the smaller gains required to reach secondary maximum of flux 
in the feature, if any (see Figure~\ref{fg}).  {These instances are quite rare. 
To estimate their contribution, we consider all features living for at least four minutes 
(eight time steps) so as to distinguish changes in flux from noise fluctuations. They 
constitute a small fraction of $\approx 4\%$. If all these features are assumed to show 
two maxima of equal strength, then they increase the FER by $\approx 
1.5\%$. This is a generous estimate and both these assumptions are unlikely to be met. 
However this is balanced out by not considering the features that have more than two 
maxima. Thus the increase in FER is quite minor.} 

Additionally, some of the features identified in a given time frame could disappear, i.e. 
drop below the noise level, for the next couple of frames, only to reappear after that. 
 {This is unlikely to happen due to the thermal or mechanical changes for the  
\sunrise{} observatory, flying in a highly stable environment at float altitude, and with 
active thermal control of critical elements in the IMaX instrument. As mentioned in 
Section~\ref{s3}, the appearance and disappearance of features could also occur due to 
the 
applied threshold on the signal levels. In our analyses, the reappeared features are 
treated as newly appeared. This leads to a higher estimation of the FER. 
\citet{2016ApJ...820...35G} have estimated that accounting for reappeared features 
decreases the FER by nearly $10\%$. If we assume the same amount of decrease in the FER 
from the reappeared features in our dataset, then we finally obtain an FER of 1100$ 
\rm{\, 
\, Mx\, cm^{-2}\,day^{-1}}$.}

\subsection{Comparison with previous studies}
\label{s6}
Below, we compare our results from \sunrise{} data with those from the \textit{Hinode} 
observations analysed in three recent publications. Although all these papers use 
observations from the same instrument, they reach very different estimates of FERs. The 
important distinguishing factor between them is the method that is used to identify the 
magnetic features and to calculate the FER. For comparison we summarize the main 
result that we have obtained here. We find that in the quiet Sun (composed dominantly of 
the IN) the FER is $1100 \rm{\,Mx\,cm^{-2}\,day^{-1}}$. This corresponds 
to $6.6 \times 10^{25} \rm{\,Mx\,day^{-1}}$ under the assumption that the whole Sun is as 
quiet as the  {very tranquil} \sunrise/IMaX FOV. 

According to \cite{2016ApJ...820...35G}, the flux appearance or emergence rate in 
the IN region is $120 \rm{\,\,Mx\,cm^{-2}\,day^{-1}}$, which corresponds to $3.7 
\times 10^{24} \rm{\,\,Mx\,day^{-1}}$ over the whole surface and the contribution from 
the 
IN is assumed to be $\approx50\%$. The authors track individual features and measure 
their 
fluxes, which is similar to the method used in LSA17. Their estimate is an order of 
magnitude lower than the FER obtained in the present paper. This difference can be 
explained by the higher spatial resolution of \sunrise{} compared to \textit{Hinode}. The 
isolated magnetic feature with the smallest flux detected in \sunrise/IMaX data is 
$9\times10^{14}$ Mx  {(see LSA17)}, which is nearly an order of magnitude smaller 
than the limit of $6.5\times10^{15}$ Mx (M. Go\v{s}i\'{c}, priv. comm.), underlying the 
analysis of \citet{2016ApJ...820...35G}. Additionally, the IMaX data are recorded with 
33\,s cadence, while the two data sets analysed by the above authors have cadences of 60 
and 90\,s each. A higher cadence helps in better tracking of the evolution of features 
and 
their fluxes.  {Also, a significant number of the very short-lived features that we 
find may have been missed by \citet{2016ApJ...820...35G}.}

\citet{2011SoPh..269...13T}, also using \textit{Hinode} observations, estimate the FER 
by fitting a power law to the distribution of frequency of emergence ($\rm{Mx^{-1} 
\,cm^{-2} \,day^{-1}}$) over a wide range of fluxes ($10^{16}-10^{22}$ Mx, which covers 
both, small-scale features as well as active regions). It is shown that a single power 
law index of -2.7 can fit the entire range. Depending on the different emergence 
detection methods used and described by these authors, such as Bipole Comparison (BC), 
Tracked Bipolar (TB) and Tracked Cluster (TC), the authors find a wide range of FERs from 
32 to $470 \rm{\,Mx \,cm^{-2}\,day^{-1}}$ which correspond to 2.0 to $28.7 \times 10^{24} 
\rm{\,Mx \,day^{-1}}$ over the whole solar surface \citep[Table 2 
of][]{2011SoPh..269...13T,th-thesis}.  To match their results from \textit{Hinode} with 
other studies, the authors choose an FER of $450\rm{\,\,Mx\,cm^{-2}\,day^{-1}}$, from the 
higher end of the range (C. Parnell, priv. comm.). This is nearly four times higher than 
the value quoted in \cite{2016ApJ...820...35G}, who also used the \textit{Hinode} 
observations and a smaller minimum flux per feature, so that they should in principle 
have caught more emerging features. However \citet{th-thesis}, using a power law 
distribution similar to \citet{2011SoPh..269...13T} and a slightly different index of
-2.5, estimates an FER of $64 \rm{\,Mx \,cm^{-2}\,day^{-1}}$.  {A possible reason 
for this difference, as briefly discussed in both these studies, could be the different 
feature tracking and identification methods used. In \citet{2011SoPh..269...13T}, all the 
features identified by BC, TB and TC methods are considered in determining the FER. 
According to the authors, the BC method counts the same feature multiple times and 
over-estimates the rate of flux emergence. However in \citet{th-thesis}, only the 
features 
tracked by TB method are used. The large differences in the FERs from the three detection 
methods quoted in Table~2 of \citet{2011SoPh..269...13T}, support this line of 
reasoning. 
 The FER in \citet{th-thesis} is roughly half that found by \citet{2016ApJ...820...35G} 
and hence is at least in qualitative agreement.} The FER estimated by us is 2.5 times 
higher than  {the largest value obtained by} \citet{2011SoPh..269...13T} and 17 
times higher than that of \citet{th-thesis}.

Another recent estimate of the FER is by \citet{2013SoPh..283..273Z}. Using 
\textit{Hinode} observations, they estimate that the IN fields contribute up to 
$3.8\times10^{26} \rm{\,Mx\,day^{-1}}$ to the solar surface. This is an order of 
magnitude higher than the global FER of $3 \times 10^{25}\rm{\,Mx\,day^{-1}}$ published 
by \citet{2011SoPh..269...13T} and is two orders of magnitude higher than the 
$3.7\times10^{24}\rm{\,Mx\,day^{-1}}$ obtained by \citet{2016ApJ...820...35G}.  In 
\citet{2013SoPh..283..273Z}, it is assumed that every three minutes, the IN features 
replenish the flux at the solar surface  with an average flux density of 
12.4\,G ($\,\rm{\,Mx\,cm^{-2}})$. Here, three minutes is taken as the average lifetime of 
the IN features \citep{2010SoPh..267...63Z}.  {Their FER is nearly six times higher 
than our estimate, although the lowest flux per feature to which \textit{Hinode}/SOT is 
sensitive is significantly larger than for \sunrise/IMaX (due to the lower spatial 
resolution of 
the former). To understand this difference, we applied the method of 
\citet{2013SoPh..283..273Z} to the \sunrise/IMaX observations. From the entire time 
series, the total sum of the flux in all features with flux $> 9\times10^{14}$ Mx is 
$1.1\times10^{21}$\,Mx over an area of $3.9\times10^{20}$\,cm$^{2}$. This gives us an 
average flux density of 2.8\,G, which is  4.5 times smaller than 12.4\,G of 
\citet{2013SoPh..283..273Z}. If the IN features are assumed to have an average lifetime 
of three minutes, similar to \citet{2013SoPh..283..273Z}, then FER over the whole solar 
surface is $8.2\times10^{25}$\,Mx\,day$^{-1}$. This is 1.2 times higher than our original 
estimate from the feature tracking method. If instead, we take the average lifetime of 
the features in our dataset from first appearance to final disappearance at the surface 
of 
$\approx1.8$ minutes, we get an FER of $1.38\times10^{26}$\,Mx\,day$^{-1}$, nearly 1.9 
times higher than our original estimate and still 2.8 times smaller than 
\citet{2013SoPh..283..273Z}. This is longer than 1.1 minute quoted in LSA17, which 
includes death of a feature by splitting or merging (see LSA17), i.e. processes that do 
not remove flux from the solar surface.}

 {To be sure that the problem does not lie in the COG technique employed here, we 
also estimated the average flux density by considering the $B_{\rm LOS}$ from the 
recently 
available inversions of the \sunrise{} data \citep{fatima}. The $B_{\rm LOS}$ values 
returned by the inversions differ from those given by the COG technique by about 5\% on 
average (individual pixels show much larger differences, of course), so that this cannot 
explain the difference  to the value adopted by \citet{2013SoPh..283..273Z}. If all the 
pixels, including noise, are considered then the average flux density is 10.7\,G. This is 
an absolute upper limit of the average flux density as a large part of it is due to noise 
 and it is still lower than the IN signal of  12.4\,G, estimated by 
\citet{2013SoPh..283..273Z}.}
%With 
%this flux density, a three minute lifetime of the IN features gives an FER of 
%$2.8\times10^{26}$\,Mx\,day$^{-1}$ and a lifetime of 2.25 minutes gives an FER of 
%$3.7\times10^{26}$\,Mx\,day$^{-1}$. 
 {Thus the high value of FER from \citet{2013SoPh..283..273Z} is at least partly 
due to their possibly too high value of average flux density. The observations analysed 
by these authors clearly show network and enhanced network features. If some of these 
are misidentified, then this would result in a higher average flux density. If this is 
indeed the case, then the estimate of the lifetime of 3 minutes may also be too short 
\citep[the technique of][neglects any possible correlation between magnetic flux and 
lifetime of a feature]{2013SoPh..283..273Z}. Although the amount of flux in IN 
fields is not expected to change significantly with time or place 
\citep[see][]{2013A&A...555A..33B}, this is not true for the amount of 
flux in the network, which changes significantly. For example another time series taken 
by \sunrise{} during its first flight, having slightly more network in the FOV, is found 
to 
have an average $B_{\rm LOS}$ of around 16\,G (including noise), which is higher than the 
12.4\,G used by \citet{2013SoPh..283..273Z}. However, this is just a qualitative 
assessment and the very large FER found by \citet{2013SoPh..283..273Z} needs to be probed 
quantitatively in a future study.}

\section{Conclusions}
\label{s8}
In this paper, we have estimated the FER in the quiet Sun from the IN features 
using the observations from \sunrise/IMaX recorded during its first science flight in 
2009. We have included the contribution from features with fluxes in the range 
$9\times10^{14} - 2.5\times10^{18}$ Mx, whose evolution was followed directly. By 
accounting for the three important processes that bring flux to the solar surface: 
 {unipolar and bipolar} appearances, and flux gained after birth by features born by 
splitting or merging over their lifetime, we estimate an FER of $1100 
\rm{\,Mx\,cm^{-2}\,day^{-1}}$. The third process is found to contribute significantly to 
the FER. The smaller features with fluxes $\le10^{16}$ Mx bring most of the flux to the 
surface. Since our studies include fluxes nearly an order of magnitude smaller than the 
smallest flux measured from the \textit{Hinode} data, our FER is also an order of 
magnitude higher when compared to studies using a similar technique 
\citep[i.e.,][]{2016ApJ...820...35G}.  {We compare also with other estimates of the 
FER in the literature. \citet{2011SoPh..269...13T} obtained a range of values. Those near 
the lower end of the range \citep[also quoted by ][]{th-thesis}, which are possibly the 
more reliable ones, are roughly consistent with the results of 
\citet{2016ApJ...820...35G}. The high FER of $3.8\times10^{26}\,{\rm Mx\, day^{-1}}$ 
found by \citet{2013SoPh..283..273Z} is, however, difficult to reconcile with any other 
study. It is likely so high partly due to the excessively large $B_{\rm LOS}$ of 
IN fields of 12.4\,G used by these authors, which is more than a factor of 4 
times larger than the averaged $B_{\rm LOS}$ of 2.8 G that we find. Even the absolute 
upper limit of the spatially averaged $B_{\rm LOS}$ in our data (including noise) is 
below the value used by \citet{2013SoPh..283..273Z}. We therefore 
expect that they have overestimated the FER.}

 {There is clearly a need for further investigation, not only to quantify  the 
reasons for the different results obtained by different techniques. There are also still 
multiple open questions. What is the cause of the increase and decrease of the flux of a 
feature during its lifetime? Is this due to interaction with ``hidden'' flux? Is this 
hidden flux not visible because it is weak and thus below the noise threshold, or because 
it is structured at very small scales, i.e. it is below the spatial resolution? How 
strongly does the ``hidden'' or missed flux change with changing spatial resolution? The 
most promising approach to answering these and related questions is to study the flux 
evolution in an magnetohydrodynamic simulation that includes a working small-scale 
turbulent dynamo.}

\begin{acknowledgements}
 {We thank F. Kahil for helping with the comparison of COG and inversion results.} 
H.N.S. and  {L.S.A.} acknowledge the financial support from the Alexander von 
Humboldt foundation. The German contribution to \sunrise{} and its reflight was funded by 
the Max Planck Foundation, the Strategic Innovations Fund of the President of the Max 
Planck Society (MPG), DLR, and private donations by supporting members of the Max Planck 
Society, which is gratefully acknowledged. The Spanish contribution was funded by the 
Ministerio de Econom\'i­a y Competitividad under Projects ESP2013-47349-C6 and 
ESP2014-56169-C6, partially using European FEDER funds. The HAO contribution was partly 
funded through NASA grant number NNX13AE95G. This work was partly supported by the BK21 
plus program through the National Research Foundation (NRF) funded by the Ministry of 
Education of Korea.
\end{acknowledgements}

\end{document}